\documentclass[aps,pre,preprint]{revtex4-1}

\usepackage{graphicx}
\usepackage{dcolumn}
\usepackage{bm}
\usepackage{physics}
\usepackage[]{siunitx}
\usepackage{textcomp}
\usepackage[mathlines]{lineno}
\usepackage{tabularx}
\usepackage{pgfplots}\pgfplotsset{compat=1.14}
\usepackage[list=true]{subcaption}

\usepackage{comment}

\DeclareSIUnit\px{pixel}
\DeclareSIUnit\k{k}

\begin{document}

\title[Coagulation paper]{Nanoparticles growth in dynamic plasma}

\author{V. Vekselman}
\affiliation{Princeton Plasma Physics Laboratory, Princeton, New Jersey 08540, USA}
\email{vekselman@pppl.gov}
\author{M. N. Shneider}
\affiliation{Department of Mechanical and Aerospace Engineering, Princeton University, Princeton, New Jersey 08540, USA}
\author{Y. Raitses}
\affiliation{Princeton Plasma Physics Laboratory, Princeton, New Jersey 08540, USA}

\date{\today}

\begin{abstract}
Coagulation growth kinetics of nanoparticles in plasma is affected by inter-particle electrostatic forces due to charging phenomenon. In stationary plasmas, unipolar charging of particles results in retardation of particles growth and may result in limitation on a particle size. We demonstrate opposite effect of enhanced particles growth in atmospheric pressure non-stationary arc discharge. Modeling of the nanoparticles growth kinetics revealed the formation of bipolar charge distribution. As a result, reversed (attractive) Coulomb forces promote formation of micrometer size particles in a millisecond time scale as observed in experiment.
\end{abstract}
\maketitle
In many different fields of nanoscale science the pathway by which particles are formed is a ctritical knowledge. In plasmas  both, intentional and unintentional production of nano-scale and micro-scale particles are commonly observed in laboratory experiments and industrial applications. These particles surrounded by a stationary plasma are subjected to charging processes resulting in formation of unipolar charge distribution as predicted by theory. As a result, Coulomb repulsion forces inhibit growth rate and limit the size of particles forming in plasma. This is observed as formation of ordered structures in plasma~\cite{Nefedov1997Quasicrystallin} characterized by topological order.
However, numerous dusty plasma studies reported about formation of large agglomerated particles thus questioning the validity of theory predictions. In attempt to explain this contradiction, models predicting attraction between similarly charges macroparticles were proposed.~\cite{Filippov2015Electrostatic-i} It was also proposed that the particles in plasma have opposite charges~\cite{Horanyi1990Coagulation-of-} and multi-group size distribution.~\cite{Shiratani1996Simultaneous-in}
A number of physical mechanisms which could potentially explain the formation of specific size and charge distributions of dusty particles were suggested including the electron emission from these particles  (secondary, thermionic, photoelectric, etc.),~\cite{Horanyi1990Coagulation-of-,Goree1994Charging-of-par,Whipple1981Potentials-of-s} charge fluctuations,~\cite{Lemons1996Scaling-laws-fo} effects of imaginary potential~\cite{Ravi2009Coagulation-of-} and ion trapping.~\cite{Goree1992Ion-trapping-by} Applicability of the orbit-motion limited (OML) theory~\cite{Mott-Smith1926The-Theory-of-C} for the description of dust-plasma interactions was addressed in Ref.~\onlinecite{Delzanno2014Charging-and-He} which developed the modified OML theory (so-called, OML+). The latter includes a more accurate description of particle charging and heat exchange processes. However, most of these models are lacking experimental validation and verification. In this work, we demonstrate a fast (sub-ms) formation of \textmu m-scale particles in a nearly thermal plasma generated by the atmospheric pressure arc discharge and propose their growth mechanism based on bipolar charging of particles. It is shown that the charging polarity of nanoparticles depends on their size.

\begin{figure}
\centering
\includegraphics[width=0.75\linewidth]{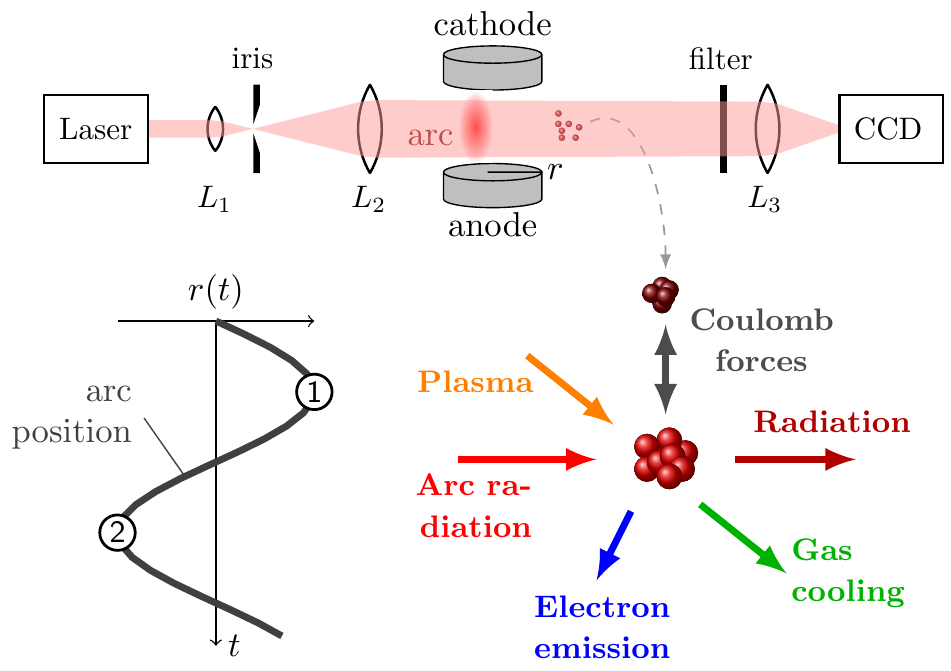}
\caption{Schematic of the experimental setup (top), arc motion (bottom left) and particle interactions (bottom right).}
\label{fig:experiment_schematics}
\end{figure}

\begin{figure}
  \includegraphics[width=0.75\linewidth]{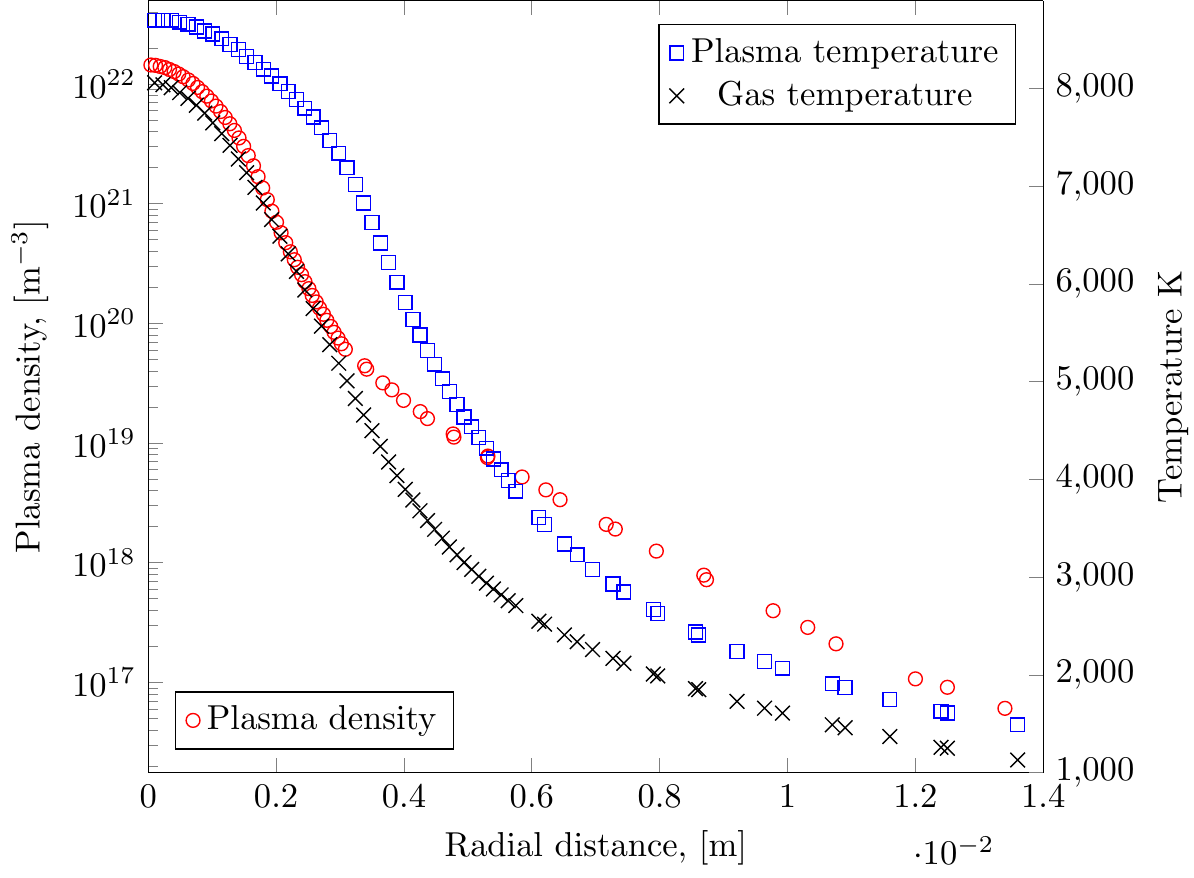}
  \caption{Radial distribution of plasma density and temperature, and gas temperature in the arc. Profiles are calculated using ANSYS CFX code.}
  \label{fig:plasma_parameters}
\end{figure}

The arc geometry and operating conditions are described in Ref.~\onlinecite{Vekselman2017Complex-structu}. The arc is formed between two graphite electrodes at sub-atmospheric pressure (\SI{66.7}{\kilo\pascal}) of helium (Fig.~\ref{fig:experiment_schematics}). In the arc core, the plasma temperature $T_{arc}$ is about \SI{8000}{\kelvin} and the carbon ablated from electrodes is presumably in a gas phase.~\cite{Vekselman2017Complex-structu} A computational fluid dynamic (CFD) code has been used to extend measured plasma parameters and gas temperature into the arc periphery region (at radial distances $\geq$5 mm from the arc core), see Figure~\ref{fig:plasma_parameters}; these CFD simulations (in ANSYS) were benchmarked with available experimental data.~\cite{Vekselman2018Quantitative-im} Lower temperature at the arc periphery promotes the condensation of carbon vapor and the formation of nanoparticles.~\cite{Yatom2017Detection-of-na} These conditions for nanoparticle formation are different from ones used in typical non-equilibrium dusty plasmas, which are usually operated at lower pressure (hundreds \si{\pascal}).~\cite{Shukla2009Colloquium:-Fun}

For monitoring of these particles, we used a narrow-band fast frame imaging to record the whole growth process. To enhance the camera sensitivity and time resolution a cw laser was used for backlighting. The source laser beam was shaped into a wide aperture collimated beam to ensure complete illumination of particles. A signal-to-noise ratio was further improved via suppression of plasma and electrode radiation by a narrow bandpass filter centered at the laser wavelength (\SI{632}{\nano\meter}). A set of frames showing formation of \textmu m-scale particles is presented in Fig.~\ref{fig:FFI} (see supplementary material~\cite{suppl_material} for the complete video file).
\begin{figure}
\centering
\includegraphics[width=0.9\linewidth]{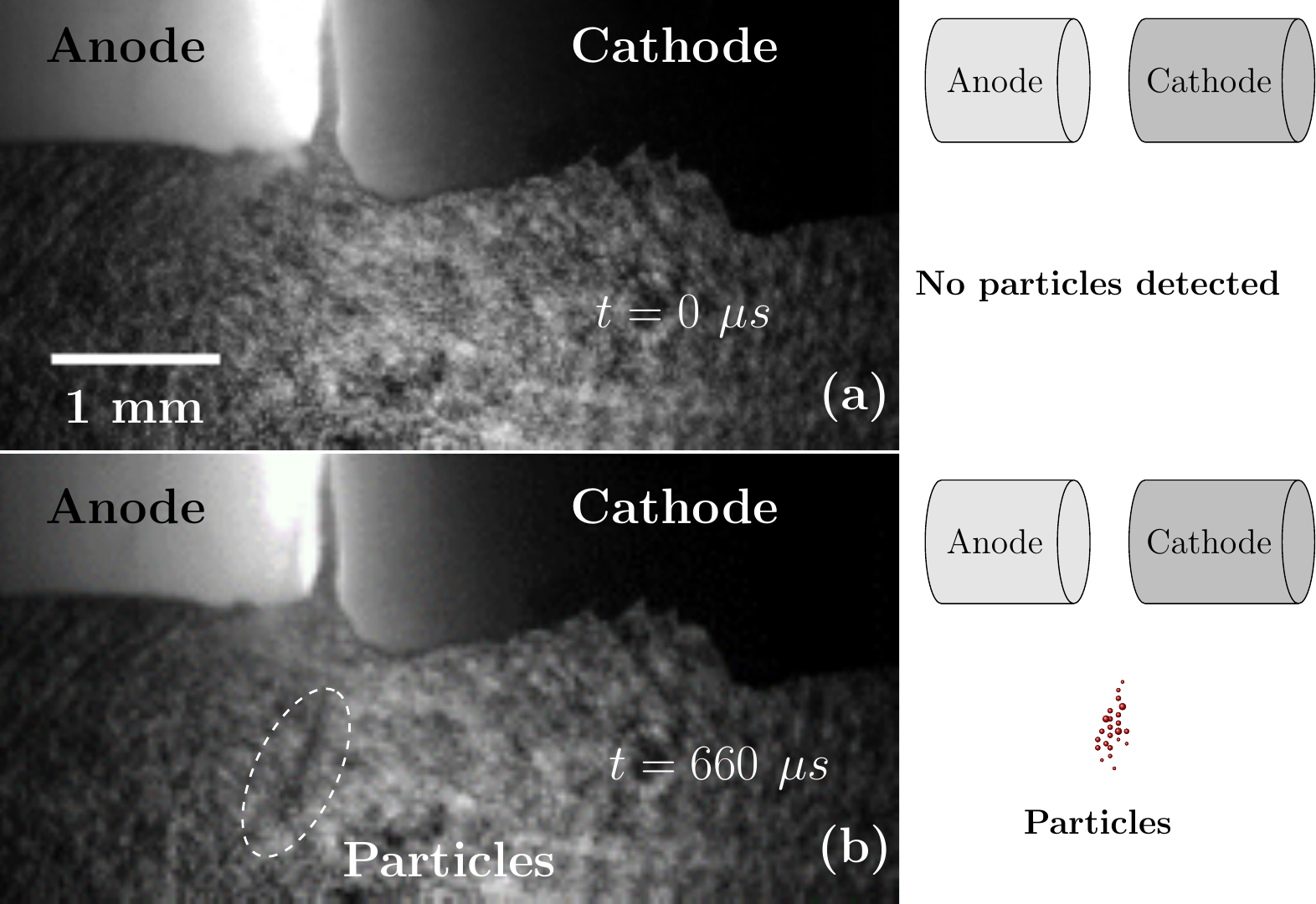}
\caption{Two selected frames (with \SI{660}{\micro\second} delay) recorded during the arc at \SI{60}{\k} fps. Video resolution is \SI{22}{\micro\meter\per\px}; frame exposure time \SI{5}{\micro\second}. Particles of \textmu m scale are encircled in the bottom frame (b) and schematically shown in the cartoon at right.}
\label{fig:FFI}
\end{figure}
The only bottom part of electrodes is captured by the camera. Uneven background is formed by laser fringes. The top surface of the anode can be partially observed in the images with a noticeable difference in the radiation intensity. The region of the highest intensity is associated with the arc attachment to the anode.~\cite{Liang2014Effect-of-arc-b} Thus, the arc is initially closer to the bottom edge of electrodes (\SI{0}{\micro\second}, Fig.~\ref{fig:FFI}a) than in \SI{660}{\micro\second} (Fig.~\ref{fig:FFI}b), which corresponds to arc location \raisebox{.5pt}{\textcircled{\raisebox{-.9pt} {1}}} and \raisebox{.5pt}{\textcircled{\raisebox{-.9pt} {2}}} in Fig.~\ref{fig:experiment_schematics}. This reflects a typical behavior of arcs demonstrating sporadic motion within an inter-electrode gap.~\cite{Gershman2016Unstable-behavi}

The most interesting feature captured at \SI{660}{\micro\second} (Fig.~\ref{fig:FFI}b) is a presence of large \textmu m scale particles. Apparently, they are formed from the gas phase and smaller size particles as evidenced by a short time interval of \SI{16.5}{\micro\second} between recorded frames (skipped in Fig.~\ref{fig:FFI}). Note that no particles were detected when the arc was away from the bottom edge of the electrodes. This observation implies a direct correlation between the proximity of the arc to the particles growth region and the formation of particles. The latter occurs at sub-ms time scale which is much shorter than a growth time typical for low pressure dusty plasmas ($\sim$ms).~\cite{Ravi2009Coagulation-of-} From these observations, we derive the following:
\begin{itemize}
\setlength{\itemsep}{-3pt}
\item[--] observed growth of particles occurs in non-stationary plasma environment;
\item[--] no external influx of particles contributes to the growth;
\item[--] Coulomb repulsion between particles is suppressed allowing the growth of particles to micrometer scale size.
\end{itemize}

In support of the above points, we propose the following explanation of the particles growth outside the arc which is complemented by the growth model described below. In the carbon arc, the electrode material is ablated and further evaporated in the hot arc core. Carbon vapor condenses in colder plasma region (arc periphery) promoting the formation of nanoparticles. For nanoparticles less than few nm in diameter, models of charge fluctuations~\cite{Lemons1996Scaling-laws-fo} and image potential~\cite{Ravi2009Coagulation-of-} describe well the process of the particle formation by coagulation in gas phase. However, larger nanoparticles are less sensitive to charge fluctuations due to accumulation of negative charge in plasma. As a result, in a steady state, the Coulomb repulsion may limit the growth of larger particles. In experiments, our imaging system can detect single micron-scale particles or clouds of smaller particles. With diffusion length of the order of few microns during the growth time, it is unlikely that large particles (Fig.~\ref{fig:FFI}b) came with the arc, but rather they formed from smaller particles.
Arc motion forms non-stationary environments for nanoparticles in the plasma. These particles are subjected to time-dependent fluxes of plasma species and heat flux. Their resulting effect on nanoparticles depends on their surface-to-volume ratio and the nature of the exerted forces. In particular, the heating of nanoparticles by the arc radiation field~\cite{Shneider2015Carbon-nanopart} should increase with the particle surface area. Then, temperature dependent thermionic emission from nanoparticles can affect the nanoparticle charging and potential with respect to the plasma.~\cite{Delzanno2004Attractive-Pote} We show below that under non-stationary arc conditions, size-dependent heating of nanoparticles and resulting thermionic emission can form a bipolar charge distribution of nanoparticles causing a Coulomb attraction and continuous growth of particles.

To mimic experimental observations when particles are exposed to time dependent charge and heat fluxes due to arc movement, we model behavior of a group of particles with diameters $\geq$10 nm placed near the arc (Fig.~\ref{fig:experiment_schematics}). We assume that at time $t=0$ the arc is located closely to the region of particles growth ($r_{min}$ at \raisebox{.5pt}{\textcircled{\raisebox{-.9pt} {1}}}) and reaches apogee ($r_{max}$ at \raisebox{.5pt}{\textcircled{\raisebox{-.9pt} {2}}}) in about \SI{1/2}{\milli\second}. We consider spherical nanoparticles of radii $r_{np}$ at distance from the arc {$r=r_{min}=$\SI{1}{\cm}} with initial temperature $T_{np}$ being equal to the local gas temperature {$T_{gas}=$\SI{1500}{\kelvin}}. The local plasma Debye length $\lambda_D$ (about 8 \textmu m for {$n_e$=\SI{1.5e17}{\per\cubic\meter}} and {$T_e$=\SI{2000}{\kelvin}}, see Figure~\ref{fig:plasma_parameters}) is comparable with the mean free path $\lambda_{mfp}$ of ion-neutral collisions $\sim$5 \textmu m between carbon ions and He atoms ($p_{gas}=$\SI{500}{Torr}). The local density of carbon neutrals is much lower that makes contribution of charge-exchange collisions negligible. At these conditions, the relation $r_{np} \ll \lambda_D \cong \lambda_{mfp}$ holds thus allowing us to apply collisionless charging model although at its very limit. We further assume the equilibrium state of plasma around nanoparticles, isotropic heating and charging, and that properties of nanoparticle material relevant to heat exchange and thermionic emission processes are being identical to larger particulates. While nanoparticles do not necessarily behave in the same manner as larger particulates in considered processes we are not aware of experimental data or modeling results which comply with this aspect and could be applied in the current study.

The plasma electron and ion fluxes to nanoparticles together with the plasma radiation lead to the heating of nanoparticles. Thermal radiation and electron emission cause cooling of nanoparticles. Moreover, there is also a heat-exchange with surrounding gas. Under dynamic equilibrium, the temperature of the particles reaches steady state when the heat fluxes to and from the particles are balanced. Since the heat exchange with the plasma is determined by the sheath potential of the nanoparticle with respect to the plasma, the steady state potential is governed by the balance of all charge fluxes between the particles and the plasma.
These heat and charge fluxes balances are described~\cite{Mitrani2016Modeling-thermi} as (see Fig.~\ref{fig:experiment_schematics})
\begin{equation}\label{eq:balance.eqs}
\begin{cases}
Mc_{heat}\frac{dT_{np}}{dt}=Q_{abs}+Q_{pl}-Q_{rad}-Q_{gas}-Q_{em},\\
C_{np}\frac{d\phi_{np}}{dt}=I_{TE}+I_{e}+I_{i},
\end{cases}
\end{equation}
where $M$, $c_{heat}$, $C_{np}$ and $\phi_{np}$ are particle mass, heat capacity, capacitance and potential with respect to plasma bulk, correspondingly. Here we have introduced
\begin{enumerate}
\setlength{\itemsep}{0pt}
\item Heat flux from radiating arc (Rayleigh regime, $\frac{r_{np}}{\lambda_{arc}}\ll 1$)~\cite{Shneider2015Carbon-nanopart}
\begin{equation*}
Q_{abs}=K_{abs}\frac{r_{np}^3 T_{arc}^5}{r^2},
\end{equation*}
where $T_{arc}$ is the arc core temperature; $K_{abs}=\frac{32\eta\pi^2 E_m\sigma_{SB}k_B}{hc}R_{arc}^2$ where $\sigma_{SB}=\frac{2}{15}\frac{\pi^4k_B^4}{h^3c^2}$ is
Stephane-Boltzmann constant, $\eta=0.8$ is emissivity of the arc; $E_m=0.35$ is a broadband value of the complex refractive index taken as for soot particles and $R_{arc}=\SI{2}{\milli\meter}$ is the arc core radius;
\item Heat flux from plasma species
\begin{equation*}\label{eq:Qplasma}
Q_{pl}=\abs{I_i}(E_{ion}-\phi_w-\phi_{np})+\abs{I_e}(\phi_w+\phi_{np}),
\end{equation*}
where $E_{ion}$ is an ionization energy for carbon (\SI{11.26}{\electronvolt}) and $\phi_w$ is the work function of the particle material (\SI{4.7}{\electronvolt}). Electron and ion currents are~\cite{Whipple1981Potentials-of-s}
\begin{equation*}\label{eq:currents}
\begin{cases}
I_k=K_k q_k (1-\frac{q_k \phi_{np}}{k_B T_k}),& \mbox{if } q_k \phi_{np}<0
\\
I_k=K_k q_k \exp{\frac{-q_k \phi_{np}}{k_B T_k}},& \mbox{if } q_k \phi_{np}>0
\end{cases}
\end{equation*}
where $k=e,i$ stands for electrons and ions, correspondingly, $K_k=n_k\sqrt[]{\frac{k_BT_k}{2\pi m_k}}$, $q_e=-e$ and $q_i=e$, $e$ is an absolute value of electron charge, $n_e(n_i)$ and $T_e(T_i)$ is plasma electron (carbon ion) density and temperature, $m_e(m_i)$ is electron (carbon ion) mass and $k_B$ is a Boltzmann constant; OML approximation is applied here ($\frac{r_{np}}{\lambda_D}\ll 1$ and $\frac{T_{np}}{T_e}\approx 1$), where $\lambda_D=\sqrt[]{\frac{\epsilon_0 k_B T_e}{2e^2 n_e}}$ is the Debye length, $T_e=T_i$);
\item Particle radiation (in Rayleigh regime)~\cite{Shneider2015Carbon-nanopart}
\begin{equation*}\label{eq:Qrad}
Q_{rad}=K_{rad} r_{np}^3 T_{np}^5,
\end{equation*}
where $K_{rad}=\frac{32\eta\pi^2 E_m\sigma_{SB}k_B}{hc}$;
\item Gas cooling~\cite{Filippov2000Energy-transfer}
\begin{equation*}\label{eq:Qgas}
Q_{gas}=K_{gas} r_{np}^2
\end{equation*}
is calculated in a free molecular regime (Knudsen number $K_n>1$); $K_{gas}=2\pi\alpha_T p\cdot \sqrt[]{\frac{R_m}{2\pi M_{He}}}\frac{\gamma+1}{\gamma-1}$ where $\alpha_T=0.1$ is a thermal accommodation coefficient for helium, $p$ and  $M_{He}$ is helium pressure and molar mass, $R_m$ is universal gas constant and $\gamma=5/3$ is specific heat ratio;
\item Cooling due to thermionic emission; here, we neglected secondary electron emission, photoemission and field emission as they are non-dominant in the arc as compared to thermionic emission. Under such conditions, the cooling is given by
\begin{equation*}\label{eq:Qthermionic}
Q_{em}=I_{TE}(\phi_w+\phi_{np}+2k_B T_{np}),
\end{equation*}
where $I_{TE}$ is the Richardson-Dushman thermoemission current
\begin{equation*}\label{eq:Ithermionic}
\begin{cases}
I_{TE}=K_{TE} \exp{\frac{-e(\phi_w-\delta\phi)}{k_B T_{np}}},& \phi_{np}<0,
\\
I_{TE}=K_{TE} (1+\frac{e\phi_{np}}{k_B T_{np}})\exp{\frac{-e(\phi_w+\phi_{np})}{k_B T_{np}}},& \phi_{np}>0,
\end{cases}
\end{equation*} %
where an exponential term with $\delta\phi=\sqrt[]{\frac{-e\phi_{np}}{4\pi\epsilon_0r_{np}}}$ accounts for the Schottky effect~\cite{Shneider2015Carbon-nanopart} and $K_{TE}=\frac{4\pi e m_e k_B^2}{h^3}r_{np}^2 T_{np}^2$.
\end{enumerate}

\begin{figure*}[th]
\centering
\begin{subfigure}[b]{0.45\textwidth}
  \includegraphics[width=3in]{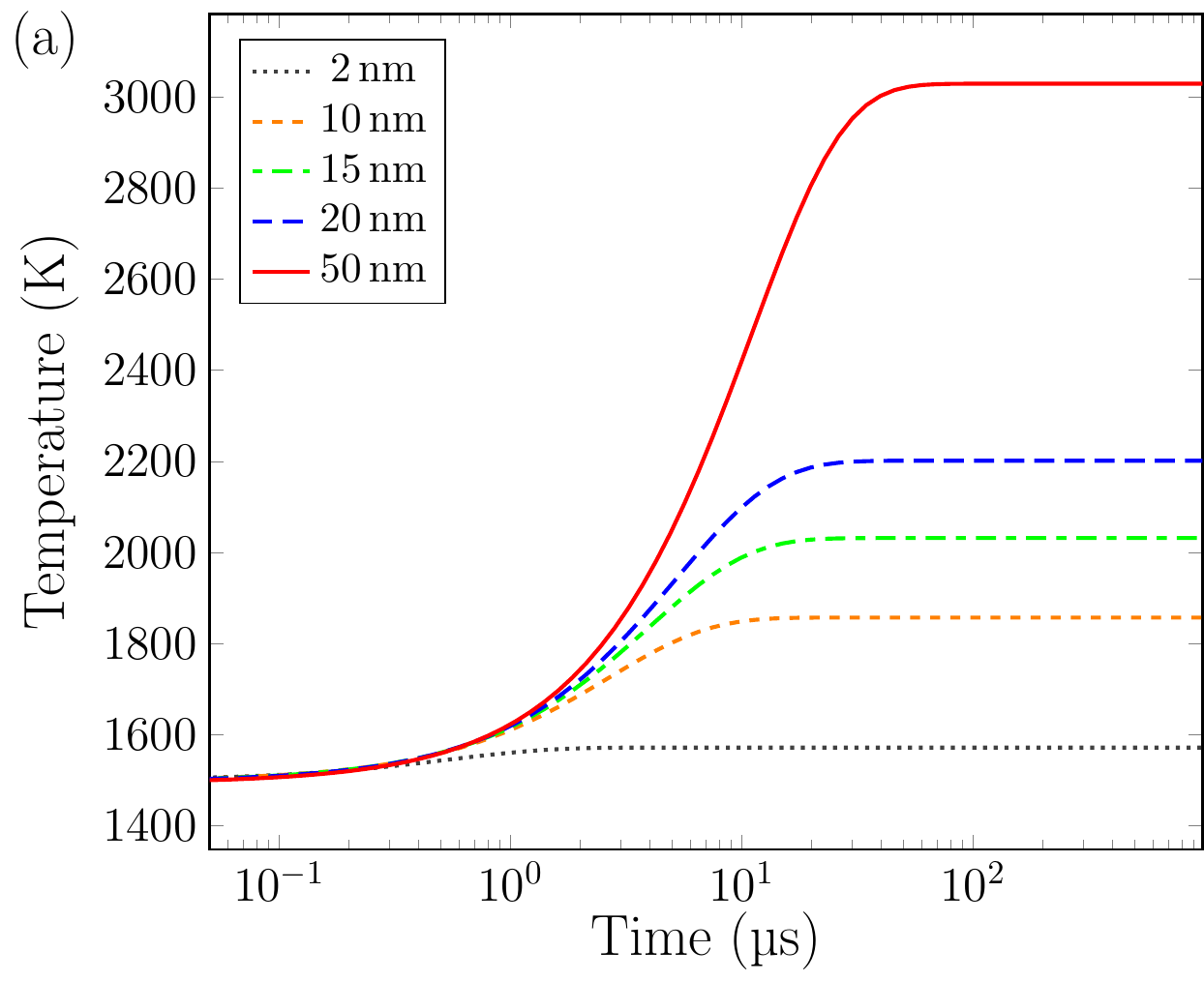}
\end{subfigure}
\begin{subfigure}[b]{0.5\textwidth}
  \includegraphics[width=3in]{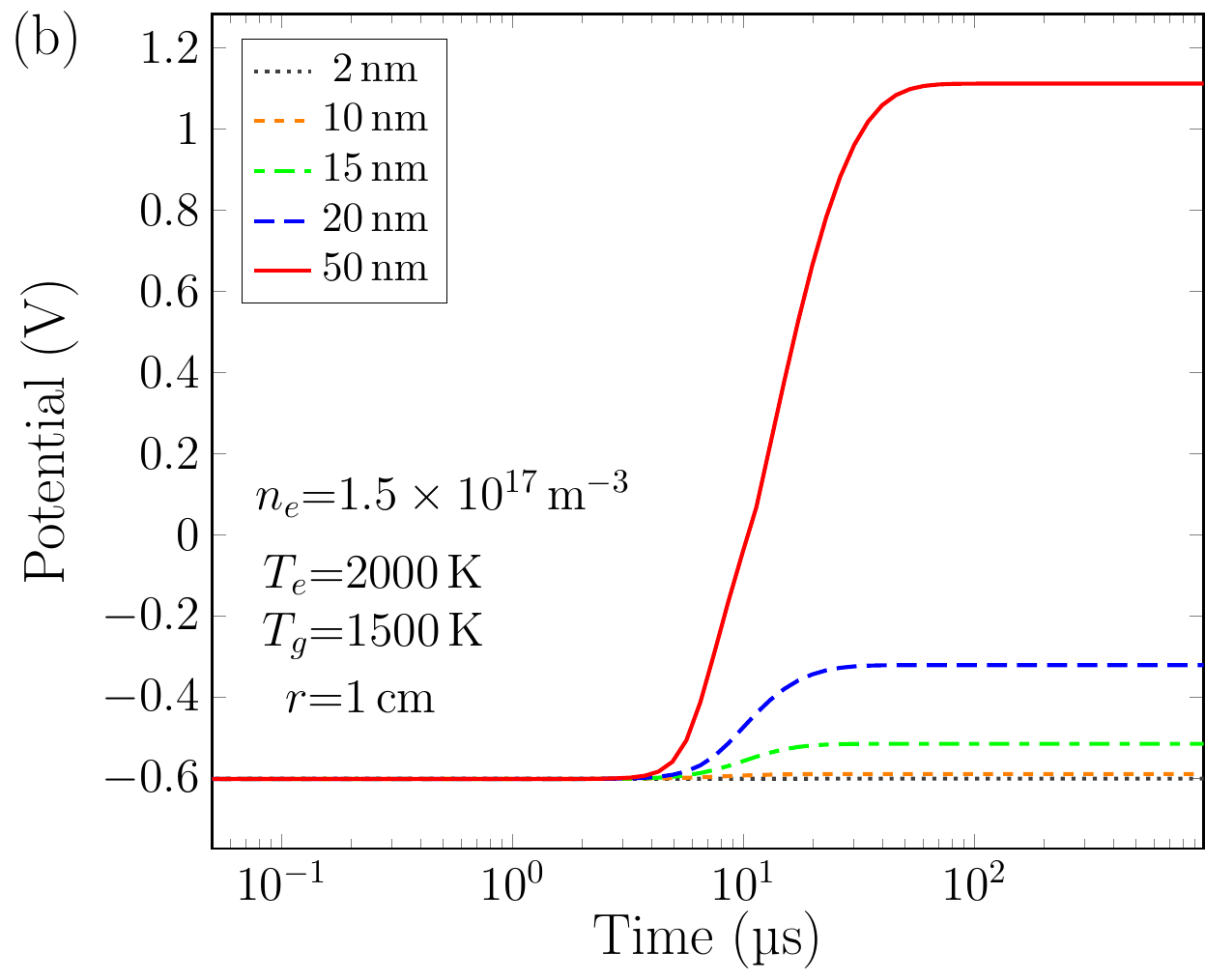}
\end{subfigure}\\
\begin{subfigure}[b]{\textwidth}
  \includegraphics[width=3in]{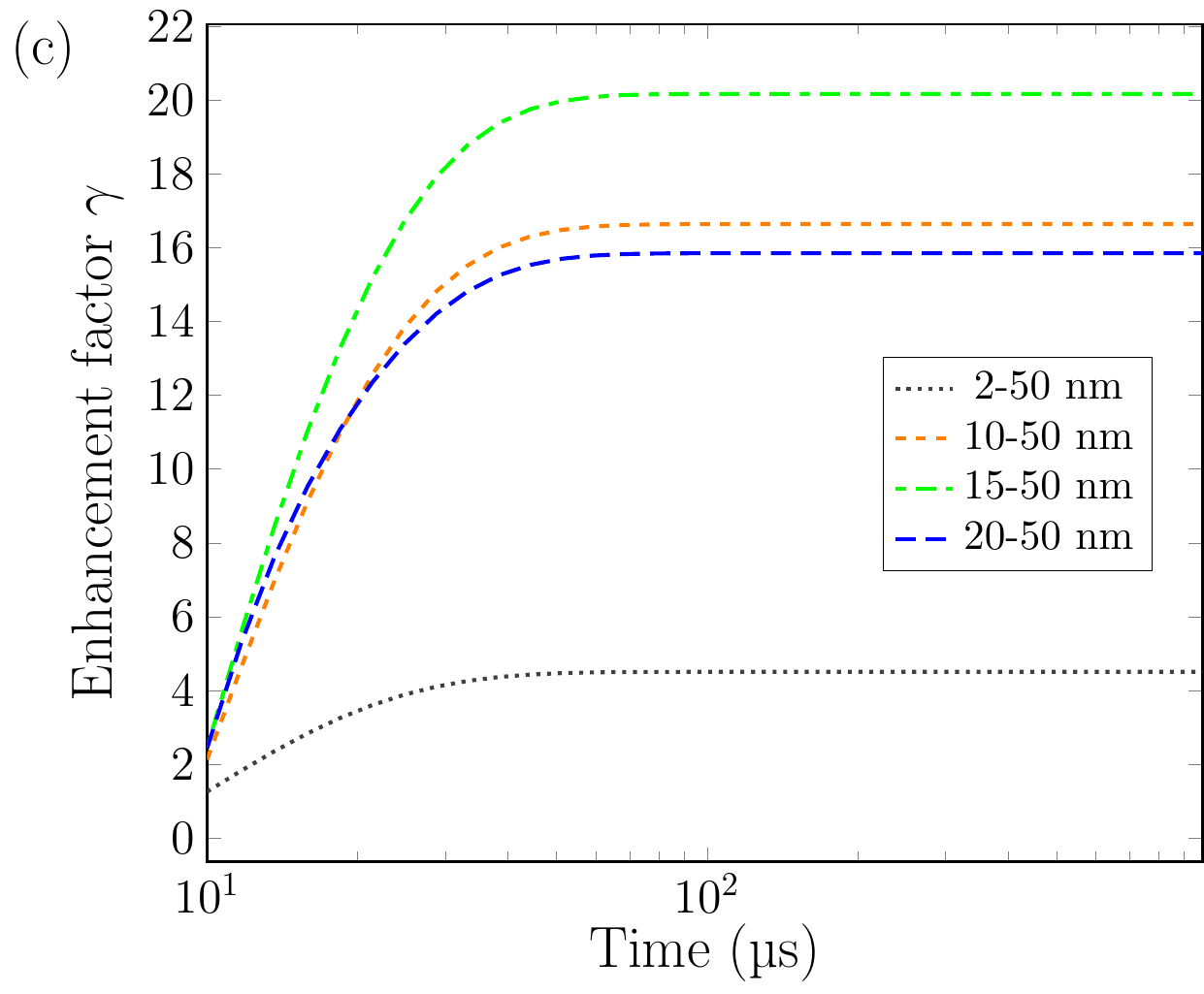}
\end{subfigure}
\caption{Time evolution of (a) temperature and (b) potential of particles and (c) enhancement factor $\gamma$ between selected pairs of particles at closest to the arc distance. Model parameters (electron density $n_e$ and temperature $T_e$, gas temperature $T_{gas}$ and distance to the arc $r=r_{min}$) used in modeling are shown in (b).}
\label{fig:heating}
\end{figure*}

Figure~\ref{fig:heating} show the time evolution of temperature and potential for considered particles by solving Eq.~\ref{eq:balance.eqs} simultaneously.
Our model predicts that for all considered sizes, the nanoparticles reach their steady state temperatures in 10-50 \si{\micro\second} (Fig.~\ref{fig:heating}a). The time to reach the equilibrium state increases with the particle size. For example, a \SI{50}{\nano\meter} particle reaches the steady-state temperature of about \SI{3000}{\kelvin} in \SI{20}{\micro\second}.
At this temperature, particle mass losses due to sublimation becomes appreciable ($\approx$3\%). For smaller particles, the temperature is lower and, as a result, the mass loss due to sublimation is estimated to be negligible ($<1\%$).
When the particle temperature is below \SI{2000}{\kelvin}, the particle charge is governed by the fluxes of ions and electrons from the plasma. This is because the thermionic emission is insignificant at such low temperatures.  As a result, the flux of emitted electrons from the particle is negligible as compared to the flux of plasma electrons to this particle. Under such conditions, the potential of the particle with respect to the plasma is at its minimum (i.e. maximum potential difference between particle and the plasma).  For the particles considered here, the minimum potential stays for a few \si{\micro\second} (Fig.~\ref{fig:heating}b). Further in time the particle temperature increases due to arc radiation absorption and saturates balancing the terms in right hand side of Eq.~\ref{eq:balance.eqs}. As a result, the thermionic emission flux increases and, depending on the particle size, can become comparable with the flux of plasma electrons or even exceeds it. This process leads to a decrease of the potential diffrence between the nanoparticle and bulk plasma as well as the number of trapped ions~\cite{Goree1992Ion-trapping-by} which can substantially affect the potential distribution near the nanoparticle. In this case the sheath potential distrubution is mostly determined by thermionically emitted electrons from the hot nanoparticle whereas the role of trapped ions is secondary.
A loss of electrons by thermionic emission coupled with undercompensated net incoming negative charge flux triggers the growth of nanoparticle positive charge. At some momemt this process equals the nanoparticle potential with the plasma potential and further makes it positive relative to the plasma. Here the potential of the nanoparticle is defined with respect to the plasma potential in the vicinity of the nanoparticle (sheath size). Having a positive potential with respect to the local plasma potential implies that the net charge of the nanoparticle is positive with respect to the space potential. This is assuming that the sheath potential changes monotonically between the nanoparticle and the plasma or any non-monotonic changes are insignificant as compared to the total potential drop in the plasma-nanoparticle sheath. In the model, the flux of charged particles (ions and electrons) is self-consistently calculated as a function of the nanoparticle potential (relative to the plasma potential) during the simulation time domain to account for these effects and carefully track the polarity change.

This process results in the formation of a bipolar size distribution of particles. In particular, for considered conditions, there is a particle size threshold ($\sim$\SI{40}{\nano\meter}) below which particles are negatively charged and above which particles are charged positive. This size-charge distribution is reversal to the distribution derived in Ref.~\onlinecite{Ravi2009Coagulation-of-} for a low pressure capacitive coupled rf plasma due to a difference in charging mechanism.

The formation of this bipolar charge distribution of particles enhances the coagulation process due to attractive Coulomb interaction. Without external forces the Brownian coagulation rate $\beta$ between neutral particles of radii $r_i$ and $r_j$ was derived by Smoluchowski. In a free molecular regime ($K_n>1$) the coagulation rate is
\begin{equation}\label{eq:beta}
\beta(r_i,r_j)=\left(\frac{3}{4\pi}\right)^{1/6}(r_i + r_j)^2\cdot\sqrt[]{\frac{6k_B}{\rho}\left(\frac{T_i}{r_i^3}+\frac{T_j}{r_j^3}\right)},
\end{equation}
where $\rho$ is a particle density and $T$ is a particle temperature. Effects of Van der Waals forces, thermophoresis, acoustic and electrostatic fields are commonly accounted via correction coefficients for the coagulation rates. Following the work of Fuchs,~\cite{Fuchs1934Uber-die-Stabil} the rate coefficient in the case of bipolar charging of particles is enhanced by a factor
\begin{equation}\label{eq:gamma}
\gamma=\frac{1-e^{-\lambda}}{\lambda}, \lambda=\frac{-\abs{q_iq_j}}{2\pi\epsilon_0(r_i+r_j)k_B(T_i+T_j)},
\end{equation}
which we further refer to as an enhancement factor; it is plotted in Fig.~\ref{fig:heating}c.

Our model shows that particles exposed to the arc radiation are subjected to bipolar charging which promotes growth rates exceeding ones between neutral particles. Furthermore, growth of big micron-size particles observed in experiment is not surpressed by the Coulomb repulsion forces.
The typical response time of particles to the variation of external conditions is about \SI{100}{\micro\second} (see Fig.~\ref{fig:heating}) and the threshold particle size at which the charge reversal happens is sensitive to the arc distancing.
It is shown that favorable conditions for continious growth of particles are naturally formed in oscillating arc (with frequencies in kHz range) as supported by observation (Fig.\ref{fig:FFI}).
It is important to emphasize that oscillating arc affecting particles manifests itself in a sporadic motion of the arc core between the arc electrodes. This motion is also a source of acoustic perturbations in the surrounding weakly ionized plasma.~\cite{Gershman2016Unstable-behavi,Popov2017Sound-produced-} As shown in Ref.~\onlinecite{Shneider2016Carbon-Nanopart}, the larger particles with a size larger than the mean free path of gas atoms/molecules ($K_n<1$) can be rapidly fused in acoustic field to micrometer size aggregates as observed in experiment. At some point, the aggregated particles become heavy enough to fall away from the growth region. This mass-separation process may limit the maximum size of particles grown in the arc. A self-consistent modeling of particle coagulation in a dynamic plasma is needed to extend this work to other laboratory and space plasmas.

In summary, sub-ms growth times of micron scale particles were observed in oscillating carbon arc at sub-atmospheric pressure. This experimental observation was modeled by accounting for time-dependent fluxes of energy and charges from plasma to the nanoparticles. Our model predicts formation of bipolar charge distribution of nanoparticles leading to the enhanced coagulation rates between oppositely charged nanoparticles. In particular, the formation of bipolar charge distribution is mainly governed by the interplay between arc-induced radiative heating of the nanoparticles and cooling of these nanoparticles by thermionic electron emission. In addition to arcs, this interplay can also be implemented in dusty plasmas with external heating of particles by for example, lasers or infrared lamps, and in plasmas generated by laser-vaporization of solid targets. Among practical applications of plasmas with a controllable bipolar distribution of particles is assembling and manufacturing of 3-D structures in plasma volume.

\section*{Supplementary Material}
This Supplementary Material~\cite{suppl_material} contains the recorded video file of particles growth in carbon arc (Fig.~\ref{fig:FFI}).

\begin{acknowledgments}
The authors are grateful to A. Merzhevsky for support with assembly of the experimental setup and to Dr. A. Khrabry, Dr. S. Yatom, Dr. I. Kaganovich and Dr. B. Stratton for fruitful discussions.

This work was supported by the US Department of Energy (DOE), Office of Science, Basic Energy Sciences, Materials Sciences and Engineering Division.
\end{acknowledgments}

\nocite{*}

\providecommand{\noopsort}[1]{}\providecommand{\singleletter}[1]{#1}%

\end{document}